# Bending of Light Near a Star and Gravitational Red/Blue Shift : Alternative Explanation Based on Refraction of Light

R.C.Gupta
Professor,
Institute of Engineering & Technology
Lucknow, India
email: rcg_iet@hotmail.com

**Abstract**

Many of the general-relativity-tests such as bending of light near a star and gravitational red/blue shift are explained without general-relativity & without Newtonian-approach. The author first casts doubts on both, the Newtonian and the relativistic approach; and proposes a novel alternative-explanation. The new alternative-explanation is based on refraction-phenomenon of optics. It predicts that as the ray passes through/near the star's atmospheric-medium, it bends due to refraction-phenomenon towards star-core, like a ray bends while passing through a prism or water-drop. A semi-empirical estimation of the atmospheric-height h and its refractive-index µ are made to find the refraction-results. The refraction-based theory also suggests new explanation for gravitational red/blue shift; it tells that frequency ? remains constant (as it is so in refraction-phenomenon) and the red/blue shift is due to change in wavelength ? due to change in velocity of light c in the medium . Estimated results for bending of light and the red/blue shift etc. with the new approach though agree well with known values, but important thing is that the physics is quite different. Also discussed are black-hole, gravitational-lensing and space-time in the new perspective of refraction. The proposed refraction-based theory proposes a new-look on black-hole, suggesting that black-hole formation is critically due to total-internal-reflection within atmosphere and subsequent absorption into the star-core. Gravitational-lensing is explained as real refraction-lensing with possibility of chromatic-aberration. The new refraction-based theory also makes a few new predictions. The present paper also suggests a possible-alternative to the Einstein's curved geometry of space-time, and indicates that the fabric of space-time which wraps(curves) around the mass is not the empty-vacuum but the atmospheric-medium. The new refraction-based approach providing alternative to general-relativity, could have important bearing on understanding of space-time, gravity and cosmology !

## 1. INTRODUCTION

As observed on Earth, light from a distant star/planet (such as Venus) bends when it passes near another star (such as Sun). Einstein's theory predicts ($4GM/c^2R$) double the bending as predicted ($2GM/c^2R$) by conventional Newtonian mechanics. Experimental confirmation was the triumph of Einstein's general theory of relativity [1,2]. In the present paper, the author, however, first casts doubts on both – the Newtonian explanation and Einstein's explanation in two subsequent paragraphs; and then proposes an alternative explanation based on refraction-phenomenon of optics. The alternative theory also explains black-hole and gravitational red/blue-shift.

Conventional Newtonian explanation for bending of light is based on photon's gravitational attraction towards the star (Sun). In fact, photon has no material mass (rest-mass zero) and has mass ($h\nu/c^2$) only due to its energy ($h\nu$). It is hypothesized that gravitation is only between material bodies; the author casts doubts on the Newtonian gravitational-attraction on photon. Though photon has energy & momentum [3]; it does not seem to have inertial & gravitational mass, else it would have been possible to accelerate or decelerate it. Moreover, if photon is considered as wave, it is not clear as how ( & if ) gravity can influence it. The author thus concludes/opines that gravitation (of sun) does not influence photon (coming from venus), therefore can not cause bending of light through Newtonian-mechanics. Also thus gravitation should not be responsible for the so-called gravitational-red/blue-shift.

Einstein's general-relativity explanation is based on geodesic or curvature of space-time near a massive body. Although general-relativity has passed several tests very well but still the real test of space-time curvature is on the way (Gravity Probe-B Relativity-Mission at Stanford University under Professor Everitt; Satellite launched on April 20, 2004, results due after 16 months in Aug. 2005). Prof. Everitt quotes [4] a letter from Thorne & Will that – "Physicist's attitudes about gravity have been heavily conditioned by general-relativity. … However, we see no reasons why Nature should confirm to the present convenience of physicists. If She has chosen to go a different route from general relativity, this will shake the foundations of physics. The Gravity Probe-B mission is an honest quest in search for truth". The truth always finds way, it takes time, however. The present author wish to mention that if Ptolemy geo-centric model can ultimately change after thirteen centuries, Newton's model can be modified after three centuries, Einstein's general-relativity theory too can also be altered in one century. All avenues for possible Truth must be kept open, even though it may seem speculative.

The author proposes alternative explanation based on refraction-phenomenon of optics, for bending of light near star. When venus-light from space (say, vacuum or near-vacuum) enters into the star's surroundings/atmosphere (comparatively denser medium) the light-ray bends towards the star (sun) due to common well known phenomenon of refraction of light. The refraction-based theory is also able to explain gravitational red/blue shift. Also, black-hole, gravitational-lensing and space-time too are considered in the new perspective. In view of the uncertainty & unavailability of information/data regarding refractive-index of atmospheric-medium and its variations; a rather semi-empirical approach, for the alternative explanation for bending of light near a star and gravitational red/blue shift etc., is appropriate and is described in the paper as follows.

## 2.  BENDING OF LIGHT NEAR A STAR : The Alternative Explanation

### 2.1 The Principle: Deviation due to refraction:

Refraction of light rays is a well known optics-phenomenon [5]. This provides an alternative explanation of bending of light near a star. When light ray, from space (near vacuum), enters the star's atmosphere (medium); the light ray bends near the star due to refraction. To illustrate the bending due to refraction, consider a spherical water-droplet as shown in Fig.1.a. When light ray enters from lighter medium (air) to denser medium (water), the droplet works as prism and thus the light-ray bends due to refraction. Similarly, when light ray enters from space-vacuum (lighter medium) to star's atmosphere (denser medium) it bends due to refraction as shown in Fig.1.b. The atmosphere extends to great heights, it becomes rarer and rarer, however; a reasonable equivalent height is shown in the figure.

The amount of bending (maximum deviation) can be estimated semi-empirically ($\delta = 2(\mu -1)$ as shown in section 2.3) as follows. Consider the limiting case when the light ray enters the atmosphere touching at point A and leaves touching at point C. The incident ray touches at $i = 90^o$ & refraction angle is r at point A and vice-versa at point C as shown in Fig.2. The angle r is thus critical angle ($\mu = 1/\sin r$), and for maximum deviation the line AC touches the star-core at point B. From the star-geometry of Fig.2, Cosec $r = R^{/}/R$ where $R^{/}$ & R are atmospheric-radius and core-radius of the star. Thus $\mu = R^{/} / R = (R + h)/R = 1 + h/R$ where h is the equivalent-height of atmosphere above the star-core (estimated in the following section 2.2). For max deviation situation (Fig.2) thus,

$$\mu = \text{Cosec } r = R^{/}/R = 1 + h/R \qquad (1)$$

### 2.2 Semi-empirical estimations of equivalent height of atmosphere h and its average refractive index μ :

Star-core is enveloped by dense & diluted gaseous surroundings (or atmosphere) with varying density & refractive-index. It is thus difficult to estimate the equivalent height h of star's atmosphere, within which properties are assumed to be uniform. Factors such as gravitation, temperature, pressure, density, radiation-pressure etc, can influence it. But the main factors for star (sun) seems to be gravitation and radiation-pressure. The dimensionless ratio h/R could probably depend (proportional) on another dimensionless quantity $GM/c^2R$ (mentioned earlier for bending of light near the star). This in a way takes into account gravitation (gravitational potential energy - GM/R) and radiation (velocity c). Thus taking (assuming) proportionality-constant or fuzz-factor as k, the star's equivalent atmospheric-height h is taken semi-empirically as:

$$h/R = k\, GM / c^2 R \quad (2)$$

In optics, refractive index of a medium $\mu = \sin i / \sin r$ is also known as ratio of velocity of light in vacuum to the medium, i.e., $\mu = c_o/c_m$. Also since velocity of light (electro-magnetic wave) $c = 1/(\varepsilon.u)^{1/2}$ where $\varepsilon$ and u are electric-permittivity and magnetic-permeability of the medium; $\mu = (\varepsilon_r u_r)^{1/2} \approx (\varepsilon_r)^{1/2}$ as relative-permeability $u_r \approx 1$, where $\varepsilon_r$ is average relative-permittivity (dielectric constant of the medium) which itself is given as $\varepsilon_r = 1 + \chi$ where $\chi$ is the average electric-susceptibility of the medium [6]. Thus $(\varepsilon_r)^{1/2} = (1+ \chi)^{1/2} \approx 1+ \chi/2$. Though $\chi$, $\varepsilon_r$ & $\mu$ vary within the atmosphere with maximum at star-core to minimum at outer-layer of atmosphere; but considering the average values of $\varepsilon_r$ & $\mu$, the average equivalent value of $\mu$ (Eq.3) and ? (Eq.4, using Eqs.1, 2 & 3) are given as,

$$\mu = 1+ \chi/2 \quad (3)$$

$$\chi = 2k\, GM / c^2 R \quad (4)$$

### 2.3 Estimation of Bending (Deviation) of Light near a Star due to Refraction-Phenomenon:

The angular deviation at entry point A (Fig.2) is (i - r), and similar deviation of the ray occurs at exit point C. So, the total deviation (bending) $\delta = 2(i - r)$. From optics consideration and using simplification & approximation, and also noting that deviation is more for higher $\mu$ & that there is no-deviation for $\mu=1$; it can be shown that deviation $(i – r) \approx (\mu -1)$. Hence the expressions for total deviation $\delta$ are given as in Eq.5a, as in Eq.5.b (using Eqs.5.a & 3) and as in Eq.5.c (using Eq.5.b & 4):

$$\delta = 2(\mu -1) \quad (5.a)$$

$$= \chi \quad (5.b)$$

$$= 2k\, GM / c^2 R \quad (5.c)$$

The total deviation (bending of light) $\delta = 2kGM/c^2R$ given by Eq.5.c is same (for fuzz factor k=2) as that predicted by the celebrated general-relativity and found experimentally correct. The approach (physics) of the present explanation, however, is altogether different and is much simpler. The new approach is based on the commonly well-known phenomenon of refraction of light; there is, however, a fuzz-factor k to account for uncertainty such as in estimation of star's atmospheric height & its refractive index. The author aims to emphasize that though refraction-phenomenon approach and general-relativity approach are in agreement as far as result is concerned but the physics of both the approaches are quite different.

### 2.4 Gravitational-Lensing (in new light as Refraction-Bending):

In perspective of refraction phenomenon discussed for bending of light, the so called gravitational-lensing [7] is in fact 'real' refraction-lensing of light due to refraction through atmospheric-layer of star or galaxy (note- both star & galaxy are surrounded with cloud of gases/materials, both can cause refraction-bending of light and thus lensing). In fact the word 'lensing' here literally means real lensing (bending of light due to refraction). But through optical-lens deviation occurs with some dispersion too, causing chromatic aberration. It is expected that here too, if the lensing is due to refraction (as said in the present paper), a little dispersion (chromatic aberration) can also occur which may possibly be found experimentally. The sky as if will look more colorful, and it is the color which will differentiate between the object & its image.

## 3. BLACK HOLE
### The New Look : Light-trapping inside due to Total-Internal-Reflection

What happens if the light ray after entering the atmosphere (at r = critical angle) suffers total-internal-reflection (due to slight change in refractive index or angle) at point C (Figs.2 & 3). The ray will thus continue to travel inside the atmosphere along a closed regular polygon as shown in Fig.3. For max possible deviation & nearness to star, the ray enters at critical-angle and touches the star-core at point B (Fig.3). If the ray AC passes above B, it will not suffer total-internal-reflection and thus will come out of atmosphere at the first-pass at point C; whereas if ray sight AC is below B the ray will be trapped (obstructed, thus absorbed) by the star-core.

For max possible deviation situation when total-internal-reflection may occur at point C as shown in Fig.3; the ray after traveling a few rounds along the polygon(s) can ultimately come-out from any vertices including C. It may be noted that refraction angle $r = 90 - \theta$ and that $2\theta \cdot n = 360$ ; thus $r = 90 - 180/n$ where n is the number of sides of the polygons shown in Fig.3. The 'minimum' possible n for a polygon to exist is 3 (triangle); so for n =3, r = 30 degrees.

The limiting case for $n = 3$, $r = 30°$ (as shown in Fig.3.c) may be looked at as if this limiting case correspond to black-hole formation. This is because if the critical angle $r < 30°$ (i.e., $\mu > 2$), the light ray will directly fall (be trapped) onto the star-core and will be absorbed, thus the ray will not come-out. So, black-hole condition is $n = 3$, $r \leq 30°$ or $\mu \geq 2$ as given in Eq.6.

$$\mu = (\varepsilon_r)^{1/2} = (1 + \chi)^{1/2} \geq 2 \tag{6.a}$$

or $\quad \chi \geq 3 \tag{6.b}$

From Eq.4 (with k=2) & Eq.6.b, the final condition for black-hole is thus given by,

$$GM/c^2R \geq \tfrac{3}{4} \tag{6.c}$$

This seems to be in reasonable (middle) agreement with the known condition $GM/c^2R = 1$ (from Newtonian red-shift approach) or $GM/c^2R = \tfrac{1}{2}$ (from general-relativity / Schwarzchild-radius), for black-hole[2,3]. The present approach also indicates that the black-hole will have a thick skirt of atmosphere with minimum $\mu = 2$ and $h = R$. Figure 3 shows: total-internal-reflection within atmosphere - as possible trapping of ray and subsequent absorption into the core, leading to formation of black-hole (for n=3; r<30°, μ>2).

The black-hole introduced here through is of new type/class (optical). It appears that the core of the black-hole is a shrunk neutron-star surrounded by thick glassy skirt of atmosphere of heavy elements (including glass-forming Si) from the remnant of the supernovae-explosion. The trapped radiation (information) can come-out as re-radiation, however differently, from the black-hole core, agreeing with Hawking [8] that 'black-holes are not so black'.

## 4. GRAVITATIONAL RED/BLUE SHIFT
### New Explanation due to Refraction-Phenomenon

According to Newtonian or Einstein's theory the gravitational red (or blue) shift $d\lambda/\lambda = GM/c^2R$ and the simple known explanation is said to be as that: when ray from space approaches towards a massive body the star or planet it gains gravitational energy, thus frequency ν increases (blue shift); but as velocity of light c assumed to be constant, λ decreases accordingly.

But the physics of the present explanation is quite different. It is considered/opined (as explained in section 1, paragraph 2) that gravitation does not influence photon; so the photon's energy hν hence ν remains constant, as is known to be so during refraction. But when from space (vacuum) the light-ray enters the atmosphere (medium) of star or planet, the velocity decreases from $c_o$ to $c_m$, ν remaining constant, hence λ decreases (blue shift). So, blue shift is explained but the reason is quite different.

Similarly, when ray goes out of atmosphere (medium) to space (vacuum) red-shift occurs. It may be emphasized that with the present explanation / theory the red ( or blue) shift is in-fact not 'gravitational red /blue shift' but 'refraction red / blue shift'.

The blue/red shift can also be estimated as follows: $d\lambda/\lambda = (\lambda_o - \lambda_m)/\lambda = (\nu\lambda_o - \nu\lambda_m)/\nu\lambda_o = 1 - c_m/c_o = 1 - 1/\mu \sim (\mu - 1)/\mu$. Since $\mu \sim 1$, and that $\mu = (\varepsilon_r)^{1/2} = (1+\chi)^{1/2} \sim 1+\chi/2$, the red/blue shift is given by Eq.7.a & 7.b and by Eq.7.c (using (Eq.7.b & Eq.4)) ;

$$d\lambda/\lambda = (\mu - 1) \qquad (7.a)$$

$$= \chi/2 \qquad (7.b)$$

$$= k\, GM/c^2 R \qquad (7.c)$$

The red/blue shift $d\lambda/\lambda$ predicted by refraction is $GM/c^2R$ from Eq.7.c with fuzz-factor k=1. This shift is in agreement with the known gravitational shift. It is not inappropriate to use some fuzz-factor as k, in view of inaccuracies/uncertainty in the model/parameters. The important thing is the physics behind the shift; the author wish to mention following points:

(i) When light enters from space (vacuum) to atmosphere (medium), there is a blue-shift; and when light goes away from atmosphere to space, there is a red-shift. This is well in agreement value-wise with Newtonian/Einstein's gravitational red/blue shift; but author would like to emphasize that physics of all the three theories (Newtonian, Einstein's and the author's refraction-based present theory) are different.

(ii) The present refraction theory of red/blue shift predicts that: once the ray goes out of atmosphere and travels farther from star, there is no more further red shift as expected for conventional gravitational red shift; the red shift only occurs when the light ray comes out of atmosphere, due to refraction phenomenon. Within the atmosphere also, there would be some red/blue shift due to variation in density or refractive-index of the medium.

(iii) When, say, for example, light enters from vacuum to atmosphere, both present and previous theories predict same result blue-shift but causes (physics) are different. Also, there is some difference in present & previous theories for what is constant & what varies. Previous (Newtonian/Einstein's) theories consider velocity of light c constant, $\nu$ increases (due to gravitational-energy/time-dilation) thus blue-shift, $\lambda$ decreases to keep c constant. Whereas, present refraction-based theory considers that energy hence $\nu$ remains constant, c decreases (from $c_o$ to $c_m$), $\lambda$ decreases (from $\lambda_o$ to $\lambda_m$) thus blue-shift. For both (present & previous) theories $\lambda$ decreases for blue-shift; but in previous theories $\nu$ increases & c remains constant whereas in present theory $\nu$ remains constant and c decreases as the ray enters into atmosphere.

## 5. <u>DISCUSSIONS AND PREDICTIONS</u>

The proposed refraction-based explanation quite successfully explains: (i) bending of light (ii) red/blue shift and (iii) other aspects such as lensing and black-hole. The author suggest the following and there could be possibilities of testing the novelties.

1. The semi-empirical estimate of equivalent atmospheric-height (h) near a 'star' is roughly $h = GM/c^2$. But this formula is no good for atmospheric-height ($h^{/}$) of 'planet or satellite', where radiation pressure is almost absent and important factors are gravity, pressure & density. For 'planet/satellite' - the formula, if any, may be entirely different from the formula for 'star'. However, looking for a similar formula and noting that velocity of sound $c_s$ is related to pressure & density; the equivalent atmospheric-height for 'planet/satellite is empirically suggested / modified roughly as $h^{/} = GM/c.c_s$ which gives reasonably possible values (order of magnitude wise) $h^{/} = 4$ Km for earth and $h^{/} = 20$ m for moon; bus as we know, $h^{/} \sim 0$ for moon for different reason (escape velocity).

2. As a daring step the author opines that gravitational-attraction is between material-bodies only. Thus gravity does not influence/attract rest-mass-less photon, or photon (electro-magnetic wave) is unaffected by gravity. Since matter-less photon doesn't has grain-mass [9], it will not have any gravitational or inertial mass either.

3. Bending of light(photon)-path is neither due to Newtonian 'gravitational-attraction' nor due to Einstein's 'geodesic', but due to refraction-phenomenon of optics within the atmosphere.

4. There should be no-refraction and thus no-bending of light around a planet or satellite with almost no-atmosphere (such as on moon).

5. Black-hole has a thick skirt of atmosphere (h = R) of high refractive index ($\mu = 2$). Black-hole physics is - first trapping of light (due to total-internal-reflection) within atmosphere and finally absorbed within the black-hole core which can re-radiate it out in due course.

6. Gravitational-lensing being the true refraction-lensing; should show some chromatic effect/aberration, which may however be too less to be noticed normally.

7. The red-shift for example, occurs only when the ray comes out of atmosphere and no further red-shift afterwards. Frequency $\nu$ remains same, wavelength $\lambda$ and velocity of light c changes. Some shift within atmosphere also possible, due to possible variation of $\mu$ within it.

8. The new refraction-based explanation is so obvious that it spares little room for doubts. In future if the potentials of this new approach is recognized/appreciated, it would possibility have important bearings on understanding of cosmology.

## 6. EINSTEIN'S AND AUTHOR'S VIEWS ON CURVATURE OF SPACE-TIME

In fact due to gravity, density thus refractive-index of the atmosphere varies in radial direction; thus during bending the light ray actually follows a curved path due to variation of $\mu$ within the medium.. This curved path (as shown by a thin free-hand drawn curved-line in Fig.1.b) is apparently considered as Einstein's 'geodesic' of general-relativity whereas in fact it is 'geodesic' due to refraction through the medium of varying $\mu$.

As per Einstein's general-relativity the 4-D empty space-time is curved (warped) around a mass. The present (author's) refraction- theory indicates that 3-D space-atmosphere may be considered curved in view of density-variation (warping) of atmosphere around the mass. With passage of 'time' as the light ray proceeds forward it follows a curved path in the 3-D 'space'-atmosphere creating an impression of 'geodesic' in 4-D space-time.

Briefly summarized (1) Einstein's view and (2) author's view on space-time and gravity are as follows :

(1) As per Einstein's general-relativity the very 'vacuum' of 4-D space-time is warped/curved around a mass; also that there is no gravity but apparently appears due to curvature of the space-time. 'Space-time curvature (warping) causes apparent-gravity'.

(2) As per author's view the gaseous-'atmosphere' in flat 4-D space-time is warped/curved around the mass; and that gravity is very much there and the warping/curving (variation) of atmospheric properties is due to gravity. 'Gravity causes variation (warping/curving) of space-time'. The space-time 'fabric' which warps (curves) around the mass is not 'vacuum' but the 'atmosphere'.

## 7. CONCLUSIONS

It is suggested that- Gravitation is only between material bodies and that the zero-rest-mass photon is unaffected by gravity. The alternative novel approach to explain phenomena such as bending of light near a star and gravitational red/blue shift is based on refraction phenomenon of optics. Bending of light is due to bending of ray due to refraction within the star's atmosphere. The red/blue shift is due to optical-phenomenon of change of wavelength (frequency remaining same) due to change in velocity of light in the atmospheric medium. Other aspects such as black-hole and gravitational-lensing are also re-examined in the new perspective of refraction-phenomenon. Interesting predictions are also made. In fact many of the general-relativity-tests are explained without general-relativity on the basis of refraction-phenomenon. The new approach could have important bearing on understanding of space-time, gravity and cosmology.

**Acknowledgement:**

The author is thankful to Dr. V. Johri, Prof. Emeritus, Dr.Balak Das, Asstt. Professor of Lucknow University and Dr. M.S. Kalara, Professor, I.I.T. Kanpur for advice & help. The author is also thankful to other Professors: Dr.Rajendra Prasad, Dr.B.D.Gupta, Dr.V.P. Gautam, Dr. Ravi Sinha, Dr. Sanjay Mishra & Rajiv Kumar for useful discussions & suggestions. Thanks are also due to Dr. D.S.Chauhan, V.C. & Dr. A.K.Khare Pro-V.C. of UP Technical University, Lucknow for encouragement and also to AICTE, New Delhi for the R & D grant. Also thanks to Veena, Ruchi, Sanjay, Chhavi, Sanjiv, Anil and Shefali for assistance.

---

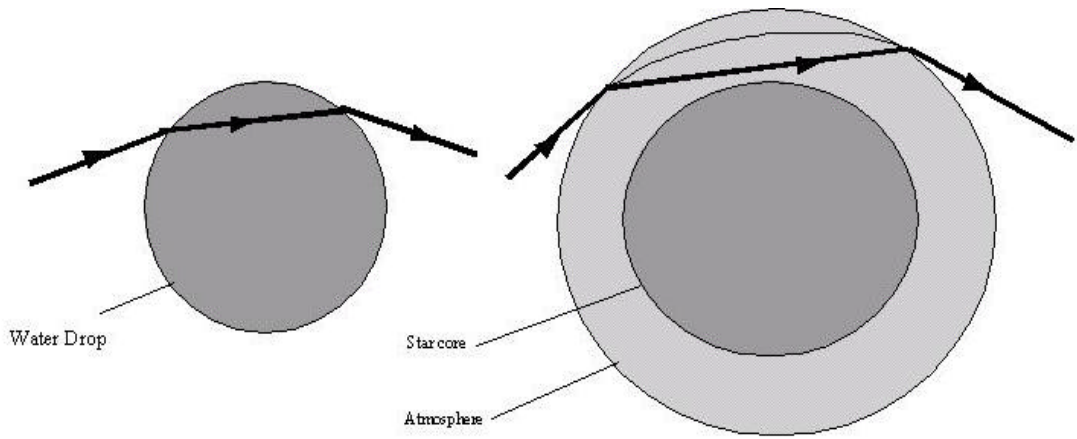

(a) Ray through water droplet

(b) Ray through atmosphere near a star

Figure 1: Refraction of light

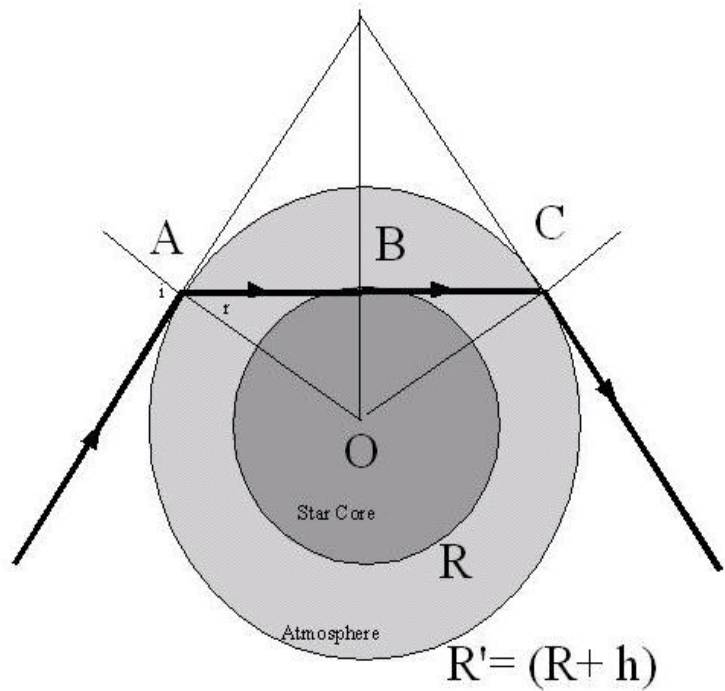

Figure 2: Bending of light near a star

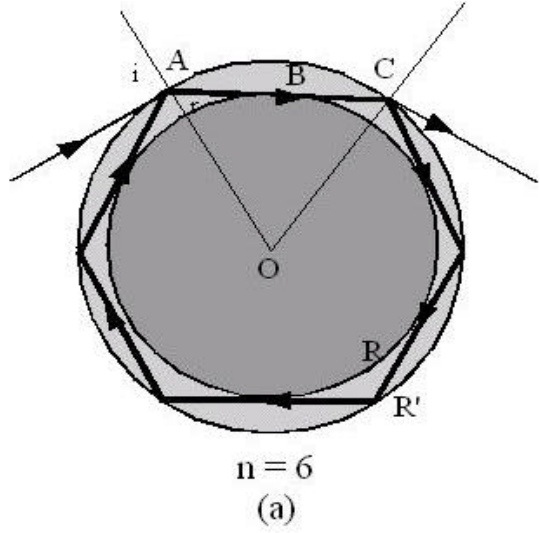

n = 6
(a)

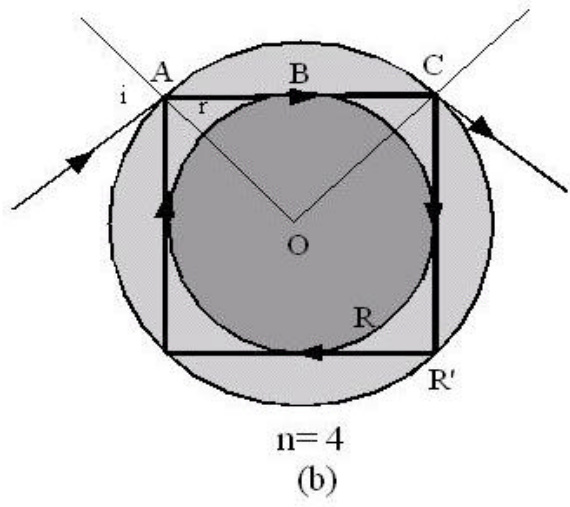

n = 4
(b)

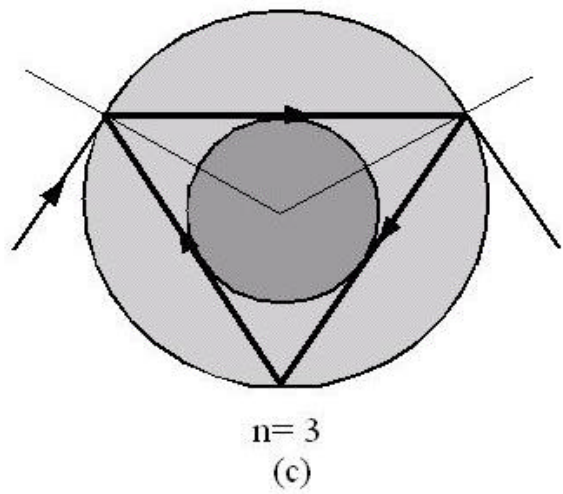

n = 3
(c)

Figure 3 : Total-internal-reflection within atmosphere as possible trapping of ray leading to black-hole (for n = 3, r < 30)